# Modeling warm dense matter formation within tight binding approximation

Nikita Medvedev[*a,b]

[a] Institute of Plasma Physics, Academy of Science of Czech Republic, Za Slovankou 1782/3, 18200 Prague 8, Czechia
[b] Institute of Physics, Academy of Science of Czech Republic, Na Slovance 1999/2, 18221 Prague 8, Czechia

## ABSTRACT

This contribution discusses challenges in modeling of formation of the warm dense matter (WDM) state in solids exposed to femtosecond X-ray free-electron laser pulses. It is based upon our previously reported code XTANT (X-ray-induced Thermal And Nonthermal Transition; N. Medvedev *et. al*, 4open **1**, 3, 2018), which combines tight binding (TB) molecular dynamics for atoms with Monte Carlo modeling of high-energy electrons and core-holes, and Boltzmann collision integrals for nonadiabatic electron-ion coupling. The current version of the code, XTANT-3, includes LCAO basis sets $sp^3$, $sp^3s^*$, and $sp^3d^5$, and can operate with both orthogonal and nonorthogonal Hamiltonians. It includes the TB parameterizations by Goodwin *et al.*, a transferrable version of Vogl's *et al.* TB, NRL, and DFTB. Considering that other modules of the code are applicable to any chemical element, this makes XTANT-3 capable of treating a large variety of materials. In order to extend it to the WDM regime, a few limitations that must be overcome are discussed here: short-range repulsion potential must be sufficiently strong; basis sets must span large enough energy space within the conduction band; dependence of the electronic scattering cross sections on the electronic and atomic temperatures and structure needs to be considered. Directions at solving these issues are outlined in this proceeding.

## INTRODUCTION

Free-electron lasers (FELs), the 4th generation of intense ultrashort XUV and X-ray radiation, opened up a new era of scientific research[1–3]. In particular, it enabled creation of the so-called warm dense matter (WDM) state routinely in a laboratory with a good control over achieved parameters[4,5]. WDM is naturally present in the universe e.g. in cores of giant planets[6]. Apart from the practical interest in this intermediate regime between conventional solids and plasma, WDM state of matter poses a fundamental interest for research bridging the gap between the respective disciplines[7].

However, a fundamental difference between the natural WDM and the transient state produced in laboratories is that the latter one is inherently nonequilibrium and short-lived[8]. FEL-induced WDM is created in multiple steps[9]: First, electronic photo-excitation by a laser pulse triggers electron cascades exciting secondary free electrons and creating nonequilibrium electronic state which typically lasts for a few tens of femtoseconds[10–12]. Simultaneously, Auger-decays of core holes occur in case if the photon energy is sufficient for their excitation. This process typically also requires a few femtoseconds (< 10 fs)[13]. Electronic excitation in some materials leads to a noticeable modification of the interatomic potential energy surface. It induces nonthermal atomic kinetics, such as nonthermal melting, or a Coulomb explosion in finite systems[14,15]. It can manifest on a scale of a few tens to hundreds of femtoseconds[16]. Typically on picosecond timescales, electron-phonon (or, more generally, electron-ion) coupling heats up the atomic system and may lead to thermal melting and ensuing damage kinetics[16]. The two processes, thermal and nonthermal, can also intertwine together[16]. Hydrodynamical expansion of the sample typically takes place at longer, nanosecond, scales[17].

To model such a large span of timescales, from sub-fs excitation to nanosecond relaxation, hybrid or multiscale approaches are required[18,19]. Such approaches, combining few models together within a unified framework with feedbacks, utilize advantages of each individual model mutually compensating for their shortcomings[20]. This allows overcoming limitations of each individual model within a hybrid scheme, thereby significantly extending the limits of applicability of the entire model. This contribution will discuss particular challenges with modeling FEL-irradiated materials on the way to WDM state.

---
[*] nikita.medvedev@fzu.cz

## XTANT CODE

The developed hybrid code XTANT[20] (which stands for X-ray-induced Thermal And Nonthermal Transitions[16]), consists of the following largely independent modules:

(a) Monte Carlo (MC) module that follows nonequilibrium high-energy electron and core-hole kinetics after an XUV or X-ray photoabsorption. It is based on asymptotic trajectory Monte Carlo scheme[21]. Photoabsorption cross sections are taken from EPICS-2017 database (formerly known as EPDL database)[22]. Compton scattering is neglected within the present version of the code, which limits its applicability to non-relativistic energies. Also, Thomson scattering of photons is not taken into account, since we are dealing with a small supercell with periodic boundary conditions, and elastic photon scattering does not play a role in this case.

The cross sections of electron inelastic scattering can be calculated within two models: (1) based on the complex-dielectric function (CDF) formalism as developed by Ritchie and Howie[23], in which case the user must provide a necessary CDF parameterization[24]; (2) within BEB atomic approximation, for which the parameters are automatically extracted from EPICS-2017 database (former EADL part of it)[22]. Elastic electron scattering for high-energy electrons (electrons with energies above a certain threshold, typically set to be 10 eV) are calculated within Mott's cross section[25].

Auger decay times of core holes are sampled with the characteristic times from the same EPICS-2017 database. Radiative decays are neglected, since for the considered parameters, such deep holes for which there would be a significant contribution of this effect are not produced[24].

(b) The module tracing a temperature evolution of valence and low-energy conduction-band electrons is using a rate equations approach. It can be regarded as a Boltzmann equation with an infinite electron-electron collision term (instantaneous thermalization), thereby enforcing a Fermi-Dirac distribution function at all times, however with an elevated temperature with respect to the atomic one[26]. The coupling of electrons to ions is calculated via the Boltzmann collision integral beyond the Fermi's Golden Rule[27].

Such a separation between the high-energy electron domain (MC) and low-energy electron domain (rate equations) is possible due to the specific shape of the transient electron distribution under FEL irradiation dubbed "bump on hot tail" distribution[12,28,29]. This transient distribution consists of nearly thermalized majority of low-energy electrons, and a long tail of high-energy electrons far from equilibrium. Hence, separate (and efficient) treatment of the two fractions with appropriate methods is justified.

(c) Molecular Dynamics (MD) model is used for tracing of the Newtonian atomic trajectories of all atoms in the super-cell with periodic boundary conditions (PBC). XTANT includes two alternative modes for PBC: constant volume simulation (V=const, NVE ensemble), or constant pressure simulation (P=const, NPH ensemble) modeled within Parrinello-Rahman scheme[30,31].

Verlet algorithm in its velocity form is used with a typical timestep of 0.01 fs for propagation of the atomic coordinates and velocities in time, as well as for the super-cell coordinates and velocities (in case of NPH ensemble)[32,33]. This timestep is chosen as to achieve convergence in calculations of both, Molecular Dynamics, as well as Boltzmann collision integral[27]. In case of Born-Oppenheimer simulations, for which the nonadiabatic energy exchange between the electrons and atoms is artificially switched off, a larger timestep can be used, which however depends on the excitation level. For WDM regime modeling when atoms can encounter steep potentials, the timestep must also be short. Additionally, a velocity scaling is used for transferring energy to atoms which was calculated from the Boltzmann integral[27].

(d) The transferable tight binding (TB) method is used calculate the transient Hamiltonian, electron band structure, collective potential energy surface for atoms, and matrix elements used in the electron-ion coupling entering the Boltzmann collision integral[27,30]. This method allows to include an influence of the electron distribution, $f_e$, on the interatomic potential, $\Phi$:

$$\Phi(\{r_{jk}(t),t\}) = \sum_{i=1}^{N} f_e(E_i)E_i(t) + E_{rep}(\{r_{jk}(t)\}). \qquad (1)$$

where energy levels are obtained by diagonalization of the orthogonal tight binding Hamiltonain: $E_i(t) = \langle i(t)|\hat{H}(\{r(t)\})|i(t)\rangle$, or by solving a secular equation (by means of Löwding orthogonalization procidure[34]) in case of a nonorthogonal one; and $E_{rep}$ is a core-core repulsive potential.

XTANT includes Slater-Koster[35] angular part of the overlap integrals for $sp^3$, $sp^3s^*$ or $sp^3d^5$ LCAO basis sets. It includes options for the radial part parameterization according to Goodwin et al.[36]; transferrable generalization of Vogl.s et al. introduced by Molteni et al.[37], NRL parameterization[38†]; and DFTB parameterization[39] by means of using Slater-Koster files[‡] (currently within the non-self-consistent TB, to be extended later).

Considering that all other modules of the code are universally applicable to any material, the bottleneck here is the availability of the TB parameterizations. With the current TB parameterizations implemented, XTANT is capable of modeling a wide variety of materials, listed in Figure 1.

(e) The complex dielectric function, which is used for calculations of optical properties, is modeled within the random phase approximation, RPA[40], which accounts for a transient band structure of the particular material without assuming free-electron gas approximation, $\varepsilon(\omega)$:

$$\varepsilon^{\alpha\beta}(\omega) = \delta_{\alpha,\beta} + \frac{e^2\hbar^2}{m^2\Omega e_0} \sum_{i,j} \frac{F_{i,j}}{(\hbar\omega_{i,j})^2} \frac{f_e(E_i) - f_e(E_j)}{\hbar\omega - \hbar\omega_{i,j} + i\hbar\gamma}, \qquad (2)$$

where $m$ is the mass of a free electron; $e_0$ is the vacuum permittivity in SI units, and $\hbar$ is the Planck's constant; $\hbar\omega_{i,j} = E_j - E_i$ is the transition energy between two eigenstates $|i\rangle$ and $|j\rangle$; $f_e(E_i)$ and $f_e(E_j)$ are the corresponding transient occupation numbers (electron distribution function mentioned above); $\omega$ is the frequency as a variable in the complex dielectric function; $F_{i,j}=|\langle i|p|j\rangle|^2$ is a part of the oscillator strength. This quantity, $F_{i,j}$, is obtained within Trani's formalism for tight binding calculations[41], generalized for nonorthogonal Hamiltonians[42].

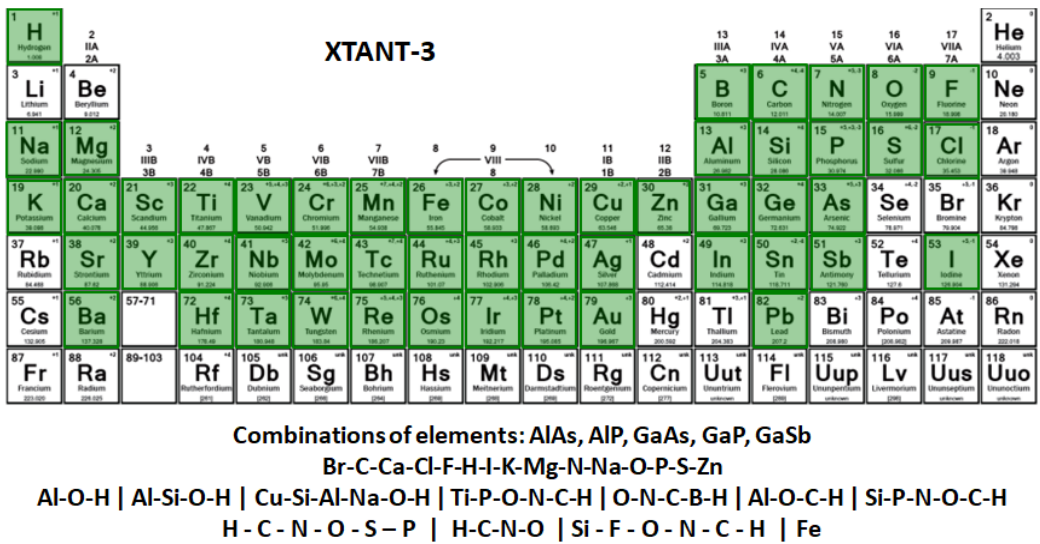

Figure 1. List of materials that can be modeled with the current version of XTANT. Elemental solids are highlighted in green within the periodic table, compounds that ban be constructed by combining elements are listed in blocks of mutually compatible parameterizations[39].

---

[†] Since some time, the NRL parameters files are no longer available at http://cst-www.nrl.navy.mil/bind/ but can still be extracted by means of internet archives such as Wayback Machine: https://archive.org/web/
[‡] See http://www.dftb.org/parameters/introduction/

## RESULTS AND DISCUSSIONS

WDM state is characterized by energies in the range of a few to a few tens of eV/atom. Thus, one has to take special care that atoms, fed with such energies, would have a proper interaction potential at short distances. For example, within Xu's *et al*. TB parameterization for carbon[36], the short-range potential only covers the region of a few eV/atom, see Figure 2. As soon as atoms have higher energies, they risk coming too close to each other and entering the divergent part of the potential, which implodes the simulations.

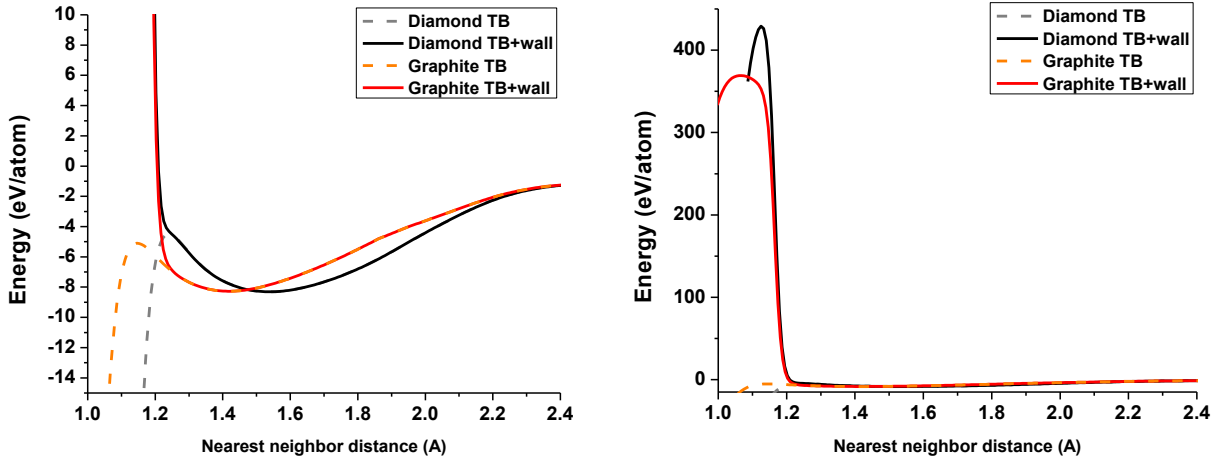

Figure 2. Comparison of the original TB potential for diamond and graphite from Ref.[36] with the additional repulsive potential at short distances (left panel), as discussed in the text. The same, zoomed out to higher energies (right panel).

The TB potential for carbon, shown in Figure 2, negatively diverges at short distances, which makes it not suitable for modeling highly excited systems, such as WDM or high pressure states. Thus, we introduced an additional part into the repulsive potential (Eq. 1), which creates an extra "exponential wall" for the atoms at short distances, $E_{EW}$:

$$E_{EW} = \frac{1}{2} \sum_{i,j}^{N_{at}} C \exp\left(\frac{1}{r - r_0}\right) f(r) \qquad (3)$$

Where the summations by *i,j* runs through all the pairs of atoms $N_{at}$, $C$ and $r_0$ are adjustable parameters, and smooth cut-off function *f(r)* is introduced as:

$$f(r) = \frac{1}{1 - \exp\left(\frac{r - r_L}{d}\right)} \qquad (2)$$

Atoms experiencing such an "exponential wall" at short distances are thus prevented from coming too close to each other. The parameters chosen for carbon are listed in Table *1*. The resulting total potential is shown in Figure 2.

Table 1. Parameters of Eqs.(3) and (4) for carbon augmenting the TB parameterization from Ref.[36].

| Coefficient | Value |
|---|---|
| **C, (eV)** | 100 |
| **$r_0$, (Å)** | 0 |
| **$r_L$, (Å)** | 1.165 |
| **d, (Å)** | 0.01 |

We must emphasize, however, that this constructed potential is not meant to precisely treat high pressure phases of carbon, and thus it should be understood rather as a numerical trick to stabilize the system; one should not expect that the high-pressure physics would be reproduced correctly. Should one need to, a dedicated fitting of parameters is required.

Let us note that other TB parameterizations do not necessarily suffer from such a shortcoming: for instance, DFTB parameterization with the spline option (but not the polynomial one) for the repulsive part includes a strong short-range repulsion in a similar manner to the one described above, and thus does not require augmenting with additional terms[43].

An example of modeling of diamond with the extended TB, including Xu's parameterization and an additional exponential wall potential Eqs.(3-4), is shown below. Diamond is numerically stable at least up to atomic temperatures of ~30 eV, as seen in Figure 3 by the good energy conservation. Here, electrons were assumed to be cold (0 K), and no electron-ion coupling was included. For such a hot atomic system, MD time step must be set small (1 as in this case).

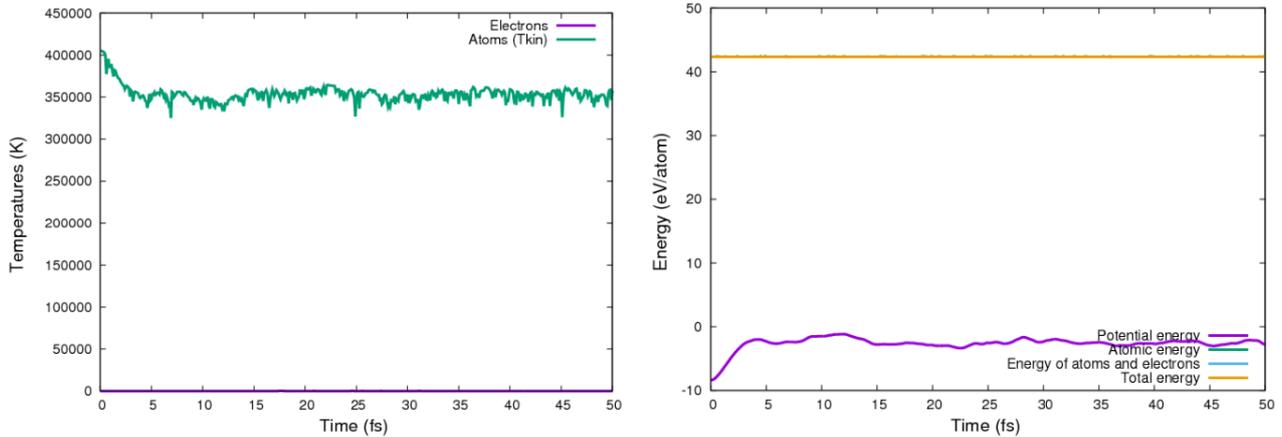

Figure 3. Temperatures (left panel) and energies (right panel) of hot diamond.

To test a full simulation scheme of XTANT, diamond was modeled under irradiation with 18.5 eV/atom absorbed dose, 6100 eV photon energy, and 6 fs FWHM pulse duration, according to a recent experiment at SACLA[44]. The energy conservation shown in Figure 4 demonstrates numerical stability of the developed scheme.

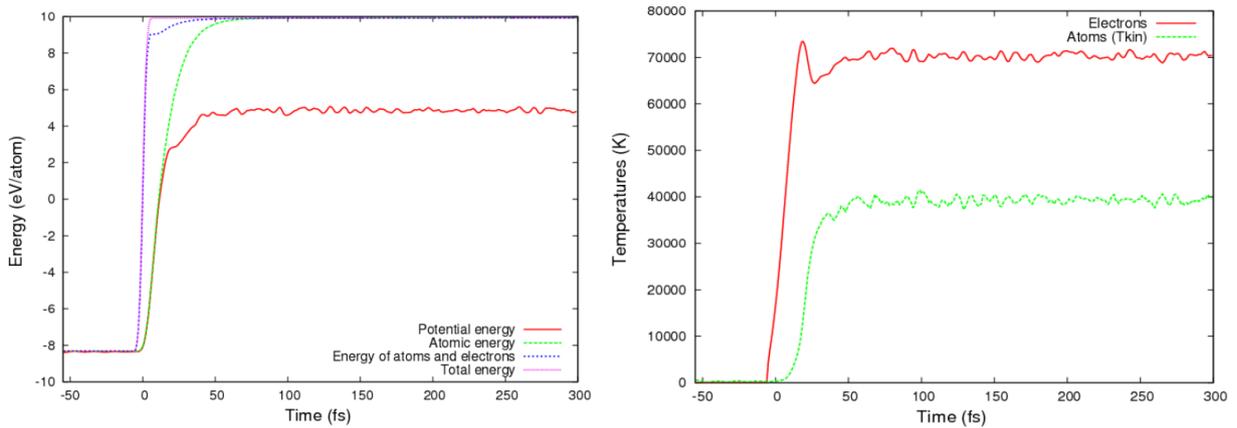

Figure 4. Energy balance in FEL irradiated diamond (left panel), and electronic and atomic temperatures (right panel).

However, such simulations in fact reach beyond the limits of applicability of XTANT for the electronic excitation. Indeed, the electron temperature reached here is very high (see Figure 4). For such an electronic temperature, the distribution function is spread to very high energies, see schematics in Figure 5. The tail of this distribution stretches beyond the top of the conduction band (CB) that can be modeled within $sp^3$-basis set from Ref.[36]. About 1% of electrons are "missing" in this case – they are artificially enforced to be within the valence band (VB) and CB, which skews the actual electron distribution function. Moreover, such a limited energy space imposes additional constraints on the energy density that electronic system can receive: above a certain threshold (in this case, around 35-40 eV/atom), the Fermi-Dirac distribution within the energy range given by a TB-basis set starts to experience divergent temperatures and chemical potentials, attempting to accommodate received energy. Limited energy space does not allow depositing an arbitrary amount of energy inside such a system.

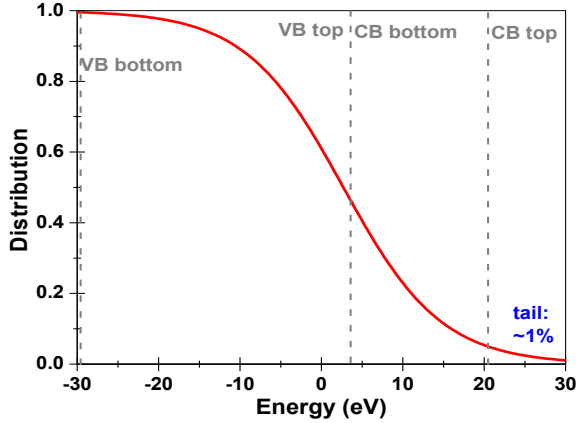

Figure 5. Electron distribution function in irradiated diamond (liquid carbon).

Currently, we have no remedy for this problem within XTANT. A straightforward approach to it could be to extend the available basis set by incorporating higher-lying electronic orbitals; to our knowledge, no such parameterization for carbon currently exists. An example of silicon is shown in Figure 6. One can see that indeed with increase of the basis set, the span of the conduction band increased. However, we must keep in mind that the TB parameters are not fitted to reproduce high-lying states within CB, and their quality there diminishes.

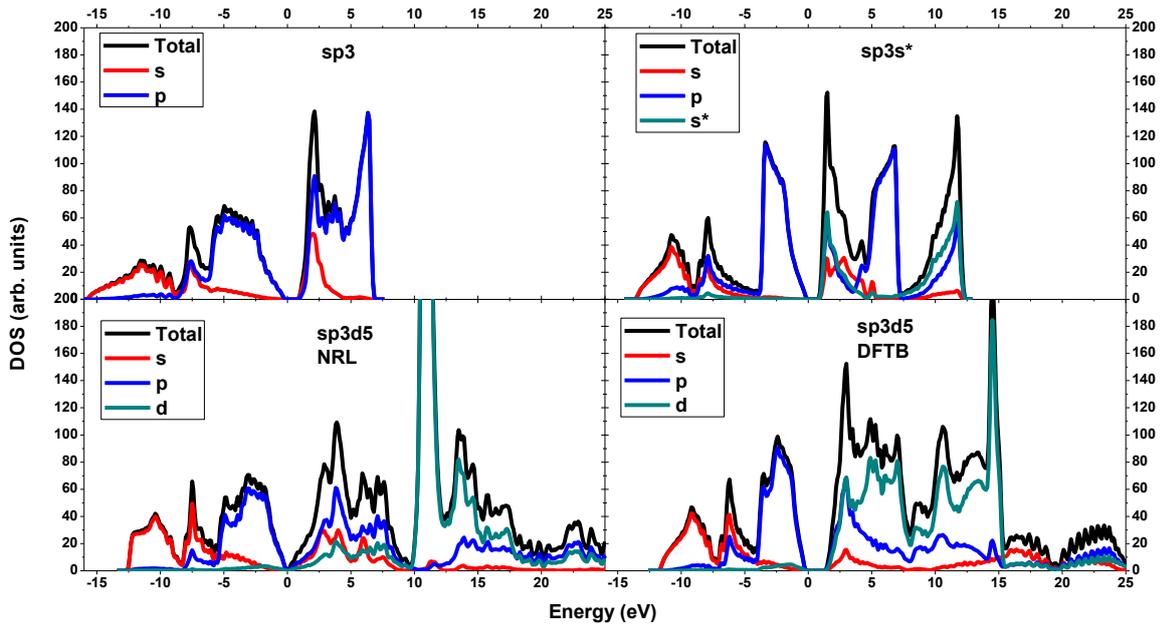

Figure 6. Partial and total density of states of crystalline silicon calculated with various TB parameterizations and basis sets: $sp^3$ from [45] (left top panel); $sp^3s^*$ from [46] (top right panel); $sp^3d^5$ from NRL[38] (bottom left panel) and from DFTB within matsci-3-0 set[39] (bottom right panel). DOSes are calculated for 216 atoms on 5x5x5 k-points grid[47]. Fermi levels are adjusted to 0 eV.

It is also possible to create TB parameterizations based not on LCAO orbitals, but on other kinds of localized basis functions, such as Gaussian functions[48]. That would also allow for a systematic extension of the basis sets. A possible simpler and computationally efficient direction of development could also be to supplement the TB-produced band structure with a free-electron band structure above the top of the TB-conduction band.

Additionally, as was mentioned in Ref.[9], the number of electrons in the conduction band is more than 1 per atom, which is far beyond the assumed low density of electrons within the CB (TB parameters are obtained for zero temperature, so they are far from their intended regime of use, and it is unclear how large of an error it introduces). The number of K-shell holes is moderate, but their effect on interatomic forces is also unknown – it has been suggested that in some cases, a presence of core holes in a solid can lead to non-thermal melting even without heating of the CB electrons[49]. These issues cannot currently be solved within TB approximation, unless parameterizations depending on the electronic temperature are constructed (we know of none).

One more difficulty with modeling WDM is with the fact that the electronic cross sections of scattering are also affected by evolution of both atomic and electronic structures. Rapid changes in the structure modify the cross sections, as discussed e.g. in Ref.[50]. It is potentially possible to account for these changes by tracing evolution of the dynamic structure factor and complex dielectric function[50] – both are available within TB – however, in this case both quantities need to be calculated with a high quality simulation, for which TB approximation might be insufficient. Extensive testing of it is required to validate the approach.

## CONCLUSIONS

In this proceeding we discussed how tight-binding-based approaches could be extended to modeling of warm dense matter states. We outlined the main challenges that one encounters by attempting to do so, and possible remedies for them. Straightforward solutions for some of them were presented, such as short-range repulsive potential extension, stabilizing the atomic systems under high energies or high pressures. Others, such as extension of the TB-produced band structure, electronic-temperature dependent TB parameterizations, and evolution of cross sections of scattering, are only mentioned as existing challenges to be aware of.

## ACKNOWLEGEMENTS

The author thanks fruitful and motivating discussions with B. Ziaja and A.E. Volkov, and H. Jeschke for bringing attention to DFTB. Partial financial support from the Czech Ministry of Education, Youth and Sports, Czech Republic (Grants LTT17015 and LM2015083) is gratefully acknowledged.